# Lattice Transformation from 2-D to Quasi 1-D and Phonon Properties of Exfoliated $ZrS_2$ and $ZrSe_2$

*Awsaf Alsulami[1], Majed Alharbi[1], Fadhel Alsaffar[1,2], Olaiyan Alolaiyan[1], Ghadeer Aljalham[1], Shahad Albawardi[1], Sarah Alsaggaf[1], Faisal Alamri[1], Thamer A. Tabbakh[3], and Moh R. Amer[1,4]\**

[1]Center of Excellence for Green Nanotechnologies,
Joint Centers of Excellence Program
King Abdulaziz City for Science and Technology, Riyadh, Saudi Arabia
[2]Department of Mechanical and Aerospace Engineering
University of California, Los Angeles, Los Angeles, CA, 90095
[3]National Center for Nanotechnology,
Materials Science Institute,
King Abdulaziz City for Science and Technology, Riyadh, Saudi Arabia
[4]Department of Electrical and Computer Engineering
University of California, Los Angeles, Los Angeles, CA, 90095

\* Please send all correspondence to <mamer@seas.ucla.edu>

**Abstract:**
Recent reports on thermal and thermoelectric properties of emerging 2-Dimensional (2D) materials have shown promising results. Yet, most of these emerging materials are underrepresented or lack a consistent and thorough investigation. Among these materials are Zirconium based chalcogenides such as Zirconium Disulfide ($ZrS_2$) and Zirconium Dieselenide ($ZrSe_2$). Here, we investigate the thermal properties of these Zirconium based materials using confocal Raman spectroscopy. We observed 2 different and distinctive Raman signatures for exfoliated $ZrX_2$ (where X = S or Se). These Raman modes generally depend on the shape of the exfoliated nanosheets, regardless of the incident laser polarization. These 2 shapes are divided into 2D- $ZrX_2$ and quasi 1D- $ZrX_2$. For 2D- $ZrX_2$, Raman modes are in alignment with those reported in literature. However, for quasi 1D-$ZrX_2$, we show that Raman modes are identical to exfoliated $ZrX_3$ nanosheets, indicating a major lattice transformation from 2D to quasi-1D. We also measure thermal properties of each resonant Raman mode for each $ZrX_2$ shape. Based on our measurements, most Raman modes exhibit a linear downshift dependence with temperature. However, for $ZrS_2$, we see an upshift (blueshift) with temperature for $A_{1g}$ mode, which is attributed to non-harmonic effects caused by dipolar coupling with IR-active modes. Moreover, the observed temperature dependence coefficient for some phonon modes of quasi 1D-$ZrX_2$ differ dramatically, which can be caused by the quasi 1D lattice. Finally, we measure phonon dynamics under optical heating for each of 2D-$ZrX_2$ and quasi 1D-$ZrX_2$ and show phonon confinement in quasi 1D-$ZrX_2$ nanosheets. We extract the thermal conductivity and the interfacial thermal conductance for each of 2D-$ZrX_2$ and quasi 1D-$ZrX_2$ nanosheets. Our calculations indicate lower interfacial thermal conductance for quasi 1D-$ZrX_2$ compared to 2D-$ZrX_2$, which can be attributed to the phonon confinement in 1D. Based on our model, we show low thermal conductivity for all $ZrX_2$ nanosheets. Our results demonstrate exceptional thermal properties for $ZrX_2$ materials, making them ideal for future thermal management strategies and thermoelectric device applications.

**Introduction:**
Since the beginning of graphene about a decade ago, new classes of 2D materials have emerged ranging from single elements such as silicene [1-3], phosphorene [4-7], and Tellurene [8-11], to Transition Metal dichalcogenides or TMDCs. These TMDCs are composed of $MX_2$ where M is a transition metal, X is a chalcogen material. Although there have been intense research efforts on TMDCs, especially, Molybdenum Disulfide ($MoS_2$) and Tungesten Diselenide ($WSe_2$) [12, 13], there are other branches of transition metal chalcogenides known as transition metal monochalcogenides (TMMCs) such as GeS, GeSe, [14, 15], and transition metal trichalcogenides (TMTCs) known as $ZrSe_3$, $ZrS_3$ [16-18]. Recent reports on these classes of materials have shown exceptional properties ranging from quantum tunneling, structural phase transitions, high thermal transport, and high thermoelectric properties [19-23].

$ZrX_2$ and $ZrX_3$ (where X = S or Se) based materials are classified as TMDCs and TMTC, respectively. Both exhibit different lattice structures, which grand them different and promising properties. For instance, $ZrS_2$ and $ZrSe_2$ exhibit high carrier mobilities reaching 1200 $cm^2V^{-1}s^{-1}$ and 2300 $cm^2V^{-1}s^{-1}$, respectively [24]. The bandgap for both $ZrS_2$ and $ZrSe_2$ is indirect with values close to 1.12eV and 1.07eV, which is close to silicon band gap [25, 26]. Such properties make $ZrX_2$ materials attractive for future device applications. On the other Hand, $ZrS_3$ and $ZrSe_3$ materials exhibit a layered anisotropic 1D superlattice chain [20]. For both of these materials, the indirect band gap energy is higher than $ZrX_2$ materials, reaching 1.88eV and 1.54eV for $ZrS_3$ and $ZrSe_3$, respectively [20, 27]. Both materials, $ZrX_2$ and $ZrX_3$, have shown exceptional thermoelectric properties with figure of merit reaching higher than 1 [27, 28].

Although there have been recent efforts to investigate the properties of $ZrX_2$ materials, there is still a noticeable gap in rigorous experimental work to understand the fundamental properties of $ZrX_2$ materials. Here, we investigate the optical properties of $ZrX_2$ materials by means of confocal Raman spectroscopy and show a lattice transformation from 2D to quasi-1D structure for $ZrS_2$ and $ZrSe_2$, evident by a radical change in the measured Raman spectra. We also investigate the temperature dependence of these Raman active modes and study the phonon dynamics for $ZrX_2$ materials. Finally, we demonstrate and discuss optothermal measurements for $ZrX_2$ materials and extract the thermal conductivity and interfacial thermal conductance for each material.

**Results:**
Figure 1a shows the lattice structure schematic of $ZrX_2$ and the primitive cell. This lattice structure exhibit trigonal shape with P-3m1 space group. Each Zr atom is connected to 6 different sulfide/selenide atoms. Atoms have a stronger bond in *ab* plane compared to *c*-axis, which makes it easier to mechanically exfoliate atomically thin $ZrX_2$. Layers are stacked and connected via a weak van der Waals force in the *c*-axis. The Raman-active modes for $ZrX_2$ materials are illustrated schematically in figure 1b. These Raman modes can be resolved spectrally as shown in figure 2a and 2b. For $ZrSe_2$, in-plane $E_g$ mode and out-of-plane $A_{1g}$ mode have frequencies at 147 $cm^{-1}$ and 194.16 $cm^{-1}$, respectively, while for $ZrS_2$, these phonon modes occur at frequencies 250 $cm^{-1}$ and 333 $cm^{-1}$, respectively. In addition, $ZrS_2$ exhibits overlapping peaks at 316 $cm^{-1}$ and 358 $cm^{-1}$. The origin of these peaks can be explained on the bases of anharmonicity effects observed for $A_{1g}$ mode where a noticeable dipolar coupling occurs between IR-active $A_{2u}$ and $E_u$ modes with $A_{1g}$ mode, giving rise to the observed broadening at $A_{1g}$ Raman shift [29]. These spectra in figures 2a and 2b have been reported previously by different groups [21, 30, 31].

In figures 2c and 2d, 3 anomalous Raman modes are detected from exfoliated $ZrSe_2$ and $ZrS_2$, respectively, which deviates greatly from typical Raman modes reported in figures 2a and 2b. These modes exhibit a Lorentzian shape with Raman shifts at 178 cm$^{-1}$, 234 cm$^{-1}$, and 302 cm$^{-1}$ for $ZrSe_2$ and 150 cm$^{-1}$, 281 cm$^{-1}$, and 320.2 cm$^{-1}$ for $ZrS_2$. Based on the optical images illustrated in figure 2c and 2d, these Raman modes only appear from exfoliated nanosheets that exhibit a noticeably narrow 1D-rectangular shape, where one axis is exceedingly longer than the other in the lateral direction (*ab* plane). In fact, due to the randomness of the exfoliation process, typical Raman modes reported in literature appear from non-rectangular (referred to as 2D-$ZrX_2$) nanosheets, while these new Raman modes in figures 2c and 2d are exclusively measured from these narrow rectangular $ZrX_2$ nanosheets (referred to as quasi 1D-$ZrX_2$).

To shed some light on the origin of these quasi 1D-$ZrX_2$ Raman modes, Raman measurements of exfoliated $ZrX_3$ materials have been measured and compared to quasi 1D-$ZrX_2$ nanosheets. In figure 3a and 3b, Raman spectra of quasi-1D $ZrS_2$ and $ZrSe_2$ are compared against Raman modes measured for exfoliated $ZrS_3$ and $ZrSe_3$, respectively. The detected Raman peaks from quasi 1D-$ZrX_2$ nanosheets remarkably overlap with the measured Raman modes of $ZrX_3$, mimicking the exact Raman spectra. Therefore, we can assign the measured peaks as $A^3_g$, $A^5_g$, and $A^6_g$ for quasi 1D-$ZrS_2$ and $A^5_g$, $A^6_g$, and $A^8_g$ for quasi 1D-$ZrSe_2$. The lattice structure and the observed phonon modes for $ZrX_3$ are schematically illustrated in figure 2c and 2d, respectively. The reason behind this identical overlapping of phonons can be attributed to lattice transformation of exfoliated 2D-$ZrX_2$ into quasi-1D structure, identical to $ZrS_3$ and $ZrSe_3$, as will be shown later.

We also measured the anisotropy behavior of 2D-$ZrX_2$ and quasi 1D-$ZrX_2$ nanosheets. In figure S1, polar plots of Raman intensity at different polarization angles are illustrated for both shapes of $ZrX_2$ nanosheets. From these plots, one can see the Raman intensity for all modes changes with changing laser polarization angle, indicating the existence of anisotropy in 2D-$ZrX_2$ and quasi 1D-$ZrX_2$ materials. This Raman intensity modulation occurs for all Raman modes with a different modulation degrees. For quasi-1D $ZrX_2$, the anisotropy trend is remarkably analogous to $ZrX_3$ materials reported previously [32], which further indicates the nature of the lattice transformation to quasi-1D structure for $ZrX_2$ rectangular shaped nanosheets.

**Temperature dependence of Raman modes:**
The Raman modes temperature dependence of 2D-$ZrX_2$ and quasi 1D-$ZrX_2$ nanosheets are measured and analyzed. Here, exfoliated nanosheets have been deposited on $SiO_2$/Si substrate. The sample is then inserted inside an enclosed temperature-controlled stage with an optical window for Raman measurements. Figure 4 shows the Raman spectra at 2 different temperatures and the change in the Raman shift vs. temperature ($\Delta\omega = \omega_T - \omega_o$) for all different $ZrX_2$ structure. For 2D-$ZrSe_2$ in figure 4a and 4b, we see an equal downshift of both Raman modes with increasing substrate temperature, mainly $E_g$ and $A_{1g}$ modes. However, for 2D-$ZrS_2$ nanosheet, a different trend is observed. $E_{1g}$ and $A_{2u}$ modes show similar downshift magnitudes with increasing temperature while $A_{1g}$ mode exhibits a weak upshift instead of a downshift, as demonstrated in the Raman spectra and Raman shift vs. temperature in figures 4c and 4d, respectively. This anomalous behavior can be attributed to the origins of $A_{1g}$ mode where anharmonicity effects are strong and caused by hyperdization between $A_{1g}$, $A_{2u}$, and acoustic phonons [30, 33]. The phonon behavior for each of 2D-$ZrSe_2$ and 2D-$ZrS_2$ has been observed in additional nanosheets, as shown in figure S2a to S2d.

We also measured Raman temperature dependence of quasi 1D-$ZrX_2$ nanosheets. In figure 4e, and 4f, we show the Raman spectra of quasi 1D-$ZrS_2$ nanosheets at 2 different temperatures and the change in the Raman shift with increasing temperatures. Although all phonon modes downshift with increasing temperature, $A^3_g$ mode shows a weak Raman dependence with temperature compared to $A^5_g$ and $A^6_g$. This weak dependence of $A^3_g$ mode stems from the quasi-1D nature of the transformed lattice of rectangular $ZrS_2$ nanosheet, where rigid interchain interaction between atoms in the 1D lattice model dominates $A^3_g$ mode, leading to this weak temperature dependence. While for $A^5_g$ and $A^6_g$ modes, we see a higher downshift with temperature indicating higher sensitivity to the surrounding temperature compared to $A^3_g$ mode. In figure S3a and S3b, other quasi 1D-$ZrS_2$ nanosheets showed analogous trend, confirming the observation of this phonon behavior.

Figure 4g and 4h show Raman spectra at 2 different temperatures and the change in the Raman shift with increasing temperatures for quasi 1D-$ZrSe_2$ nanosheet. Here, $A^5_g$, $A^6_g$, and $A^8_g$ Raman modes downshift with increasing temperature. However, it is observed that $A^6_g$ and $A^8_g$ modes exhibit higher downshift compared to $A^5_g$ mode, leading to a higher temperature sensitivity. Figure S3c and S3d also demonstrates this trend on another quasi 1D-$ZrSe_2$ nanosheet.

To quantify the observed Raman downshift with temperature for each measured Raman peak, the Raman temperature dependence of each mode can be expressed as:

$$\omega(T) = \omega_o + v_T T \qquad \text{Eq. 1}$$

Where $\omega$ is the Raman shift in cm$^{-1}$, $\omega_o$ is the Raman shift when the temperature goes to absolute zero, and $v_T$ is the temperature coefficient of the Raman mode of interest. Using this relation, one can extract the temperature coefficient for all $ZrX_2$ nanosheets. Figure S4 shows independent plots for each Raman mode measured with changing temperature. Accordingly, the obtained temperature coefficient ($v_T$) and Raman shift at 0K ($\omega_o$) for each Raman mode are summarized in table 1. We can deduce from this table that $E_g$ and $A_{1g}$ modes show similar downshift for 2D-ZrSe$_2$. While for 2D-ZrS$_2$, A$_{1g}$ show a weak and anomalous upshift with temperature compared to E$_g$ and A$_{2u}$ modes.

As mentioned above, for quasi 1D-ZrSe$_2$, all phonon modes downshift. However, the temperature coefficient for A$^8_g$ is higher than other Raman modes confirming this high temperature sensitivity. Similarly, the temperature coefficient of A$^5_g$ and A$^6_g$ for quasi 1D-ZrS$_2$ show almost equal values, suggesting equal downshift with temperature. Nevertheless, the temperature coefficient of A$^3_g$ mode shows 3 orders of magnitude less than A$^5_g$ and A$^6_g$, confirming this weak downshift with temperature as explained above.

Table 1: Raman shift ($\omega_o$) and the temperature coefficient ($\nu_T$) of each Raman mode for 2D-ZrX$_2$ and quasi 1D-ZrX$_2$

| Material | Raman mode | $\omega_o$ (cm$^{-1}$) | $\nu_T$ (cm$^{-1}$/K) |
| --- | --- | --- | --- |
| 2D-ZrSe$_2$ | E$_g$ | 146.4209 | -0.01343 |
|  | A$_{1g}$ | 194.58626 | -0.01282 |
| 2D-ZrS$_2$ | E$_g$ | 250.09365 | -0.01869 |
|  | A$_{1g}$ | 333.13913 | 0.00589 |
|  | A$_{2u}$ | 317.27941 | -0.0201 |
| Quasi 1D-ZrSe$_2$ | A$^5_g$ | 178.17638 | -0.01153 |
|  | A$^6_g$ | 235.24741 | -0.01133 |
|  | A$^8_g$ | 302.72557 | -0.01558 |
| Quasi 1D-ZrS$_2$ | A$^3_g$ | 150.48667 | -0.00768 |
|  | A$^5_g$ | 280.72034 | -0.02188 |
|  | A$^6_g$ | 319.56597 | -0.02046 |

**Localized Laser Heating of Raman modes:**

To gain a deeper understanding of the phonon dynamics of ZrX$_2$-based materials with different shapes, localized laser-induced optothermal experiment is carried out on exfoliated ZrX$_2$ nanosheets. In this experiment, a laser source is used to induce localized heat on the targeted ZrX$_2$ nanosheets, while Raman spectra are detected and measured simultaneously. It is possible to extract the temperature of a specific Raman mode with increasing laser power by obtaining the empirical relationship between this Raman mode and temperature based on equation 1. This method has been applied previously to extract the thermal conductivity of various 2D materials including graphene, MoS$_2$, and MoSe$_2$ [34-36].

Figure 5 shows the optothermal measurements obtained for 2D-ZrX$_2$ and quasi 1D-ZrX$_2$ nanosheets. For 2D-ZrSe$_2$ and 2D-ZrS$_2$ nanosheets, A$_{1g}$ and A$_{2u}$ modes were chosen to extract nanosheet temperature, respectively. Figures 5a,b and 5c,d show the Raman spectra at 2 different laser powers and the extracted temperature vs. laser power for each 2D nanosheet. A monotonic increase in temperature is observed with temperatures reaching as high as 171°C for ZrSe$_2$ and 282 °C for ZrS$_2$.

In contrast, figures 5e and 5g show the Raman spectra at 2 different laser powers for quasi 1D-ZrX$_2$ nanosheets. Here, A$^8_g$ mode and A$^6_g$ mode are chosen for quasi 1D-ZrSe$_2$ and quasi 1D-ZrS$_2$, respectively, due to their temperature sensitivity as demonstrated in the Raman temperature dependence experiment reported above. Figures 5f and 5h show the extracted temperature at each corresponding laser power. Analogous to 2D-ZrX$_2$ materials, the observed temperature trend is monotonic. However, the maximum temperature reached for the same delivered optical energy is different where quasi 1D-ZrSe$_2$ nanosheet reaches a temperature of 140 °C, while quasi 1D-ZrS$_2$ reaches a temperature close to 200 °C. These temperature values are lower than the maximum temperature observed for 2D-ZrX$_2$ nanosheets. This lower maximum temperature observed in quasi 1D-ZrX$_2$ nanosheets compared to 2D-ZrX2 nanosheets can be attributed to the lower dimensionality of the lattice where phonon confinement in 1D is pronounced compared to 2D lattice structure. This behavior will be elaborated more in the discussion section below.

**DISCUSSION:**

The observation of A$^3_g$, A$^5_g$, and A$^6_g$ for quasi 1D-ZrS$_2$ and A$^5_g$, A$^6_g$, and A$^8_g$ for quasi 1D-ZrSe$_2$ suggests lattice transformation from ZrX$_2$ to ZrX$_3$. Considering lattice reconstruction from 2D structure to 1D structure, it is possible to obtain these 3 distinct Raman modes for ZrX$_2$. During mechanical exfoliation, when quasi-1D nanosheet structures are obtained, it is possible that atoms rearrange themselves to favor the formation of ZrX$_3$ 1-D lattice structure. In fact, based on DFT calculations, the formation energy for ZrS$_3$ and ZrSe$_3$ is less than the formation energy of ZrS$_2$ and ZrSe$_2$ by 0.29eV/atom and 0.28eV/atom, respectively [37-42]. This lower energy enables atoms to favor the formation of ZrX$_3$ instead of ZrX$_2$ for quasi-1D structures, leading to lattice transformation from octahedral to trigonal prismatic chain lattice. Therefore, it is believed that quasi 1D-ZrX$_2$ nanosheets transform into ZrX$_3$ lattice structure when these 1D nanosheets are exfoliated, evident by the identical resonant phonons observed in intrinsic ZrX$_3$ nanosheets. This phenomenon can unlock the path to engineer ZrX$_3$ nanostructures from ZrX$_2$ crystals through deterministic control of mechanical exfoliation parameters.

Considering a qualitative 1D lattice model reported previously [43], one can explain the discrepancy in the weak downshift of A$^3_g$ Raman mode with varying temperature in quasi 1D-ZrS$_2$ compared to A$^5_g$ and A$^6_g$. Here, A$^3_g$ mode is affected by the interchain interaction of atoms and is marginally affected by the intrachain interactions. In contrast, A$^5_g$ is affected by both, interchain and intrachain interactions while A$^6_g$ mode is affected by the intrachain interaction of atoms. Previous reports have shown that interchain interaction is much weaker than intrachain interaction [43]. Accordingly, it is believed that the downshift of A$^3_g$ mode should differ from A$^5_g$ and A$^6_g$ modes as shown in figure 4h. Indeed, this phonon trend is also observed in ZrS$_3$ as shown in figure S5a,d. In addition, for quasi 1D-ZrSe$_2$, the phonons temperature dependence also mimics ZrSe$_3$, as shown in figure S5c,d, further cooperating this lattice transformation.

It is not fully clear why the anomalous upshift (cooling) of $A_{1g}$ mode in 2D-$ZrS_2$ with temperature. However, anharmonicity of $ZrS_2$ lattice can play a major role in this upshift with temperature. Recent calculations on anharmonicity of $ZrS_2$ phonons using self-consistent phonons reveal a pronounced anharmonicity effect with increasing temperature [44], which evidently can alter phonon dynamics with temperature. In addition, the dipolar coupling between Raman-active $A_{1g}$ phonon mode and infrared-active phonons ($A_{2u}$ (LO) and $E_u$ modes) due to the long-range coulomb forces caused by charges on atoms can contribute to the observation of this anomalous upshift of $A_{1g}$ mode and downshift of $A_{2u}$ mode [21, 29, 33, 45]. Therefore, additional investigations into the origins of this upshift should be sought in the future.

The observation of lower maximum temperature for quasi 1D-$ZrX_2$ compared to 2D-$ZrX_2$ nanosheets in figure 5 can be attributed to phonon confinement in 1D, where phonon transport is restricted only in one direction due to reduced dimensionality. In fact, evidence of this phonon confinement in 1D lattice can also be observed when two different optothermal energies are delivered, as demonstrated in figure S6. In this case, Raman spectra at 2 different optothermal energies show the same exact downshift for quasi 1D-$ZrX_2$ nanosheets. In contrast and as expected, 2D-$ZrX_2$ nanosheets show a higher downshift with increasing optical energy, as shown in figures S6a and S6b. Accordingly, it is expected that $ZrX_3$ nanosheet should exhibit similar trend to quasi 1D-$ZrX_2$. Indeed, in figure S7, we show identical behavior for $ZrX_3$ nanosheet where Raman modes show a saturated downshift regardless of the optothermal energy delivered. It is readily concluded that phonon confinement in 1D lattice plays a major role behind the observed lower downshift (temperature) of quasi 1D-$ZrX_2$ nanosheets compared to 2D-$ZrX_2$ nanosheets.

The thermal conductivity of 2D-$ZrX_2$ and quasi 1D-$ZrX_2$ (or $ZrX_3$) can be extracted using the results obtained from optothermal measurements. This method has been applied in previous reports [34, 36]. Here, assuming a diffusive heat transport, the heat equation can be described in the radial direction ($r$) according to:

$$\frac{Q}{\kappa} = \frac{g}{\kappa t}(T(r) - T_a) - \frac{1}{r}\frac{d}{dr}\left(\frac{edT(r)}{dr}\right), \qquad \text{Eq. 2}$$

where $Q$ is the optical heating (per volume) which depends on the laser power ($P$), the laser diameter ($r_o$), the nanosheet absorption coefficient ($\alpha$) and thickness ($t$) which can be obtained from AFM images in figures S8 and S9. $Q$ is given by $Q = \frac{P\alpha}{\pi r_o^2 t}\exp(-r^2/r_o^2)$. $\kappa$ is the thermal conductivity of the material, $g$ is the interfacial thermal conductance per unit area, $T(r)$ is the temperature profile as a function the center of the laser beam (assuming a gaussian laser profile), and $T_a$ is ambient temperature and is taken as 300K. Here, the measured temperature rise ($\Delta T_m$) using laser-induced Raman spectroscopy method is given by:

$$\Delta T_m = \frac{\int_0^\infty \Delta T(r) r e^{-r^2/r_0^2} dr}{\int_0^\infty r e^{-r^2/r_0^2} dr}, \qquad \text{Eq. 3}$$

Where the measured temperature is given by $T_m = T_a + \Delta T_m$. By solving this differential equation, it is possible to extract the thermal conductivity for each measured nanosheet. Table 2 shows the extracted thermal conductivities and interfacial thermal conductance using this model. The extracted thermal conductivity values for 2D-ZrX$_2$ are remarkably in agreement with theoretically predicted values [22, 46-48]. Moreover, the thermal conductivity obtained for each quasi 1D-ZrX$_2$ is also in good agreement with ZrX$_3$, further cooperating the lattice transformation into 1D lattice structure. Our fits also show the interfacial thermal conductance is lower for quasi 1D-ZrX$_2$ materials compared to 2D-ZrX$_2$ nanosheets, which can stem from phonon confinement in 1D caused by lower lattice dimensionality to 1D compared to 2D-ZrX$_2$.

Table 2: extracted thermal conductivity and interfacial thermal conductance of ZrX2 materials

| Material | Thermal Conductivity ($\kappa$) (W/m.k) | Interfacial Thermal Conductance ($g$) (MW/m$^2$ k) |
|---|---|---|
| 1D-ZrS$_2$ | 41.75 ± 21 | 6.1 ± 3 |
| 1D-ZrSe$_2$ | 17.78 ± 9 | 3.12 ± 1.5 |
| 2D-ZrS$_2$ (ZrS$_3$) | 18.8 ± 9 | 17.5 ± 8 |
| 2D-ZrSe$_2$ (ZrSe$_3$) | 28.6 ± 14 | 20.44 ± 10 |

**Conclusion:**

In summary, phonon properties of exfoliated ZrX$_2$ materials have been investigated. Two distinctive Raman signatures were detected from exfoliated ZrX$_2$ nanosheets. Arbitrary 2D shaped ZrX$_2$ referred to as "2D-ZrX$_2$" exhibit phonon modes that are in good agreement with those reported in literature. However, rectangular shaped ZrX$_2$ referred to as "quasi 1D-ZrX2" exhibit Raman signatures that are identical to Raman spectra of ZrX$_3$, where Lorentzian Raman modes are detected. These observed Raman spectra from 1D-ZrX$_2$ suggests lattice transformation from 2D to quasi-1D, giving rise to these resonant Raman modes.

Moreover, temperature dependence of each measured Raman mode is demonstrated. For 2D-ZrS$_2$, an anomalous upshift occurs for $A_{1g}$ mode which is attributed to anharmonicity of this phonon mode due to coupling with IR-active modes. For quasi 1D-ZrS$_2$, $A^3_g$ mode shows very weak Raman shift with temperature, which is attributed to the quasi-1D nature of the transformed lattice, where rigid interchain interaction between atoms in the 1D lattice model dominates $A^3_g$ mode, leading to this weak temperature dependence.

We also report laser-induced optothermal measurements of exfoliated $ZrX_2$ materials and show that 2D-$ZrX_2$ materials exhibit higher maximum temperature than quasi 1D-$ZrX_2$ materials. This behavior is attributed to phonon confinement in 1D for quasi 1D-$ZrX_2$ materials compared to 2D-$ZrX_2$ materials, which is also supported by temperature saturation when the delivered optothermal energy is increased.

Finally, we extract the thermal conductivity and interfacial thermal conductance of each material and show that the interfacial thermal conductance values of 1D-$ZrX_2$ (and $ZrX_3$) are much lower than 2D-$ZrX_2$, further cooperating the quasi-1D nature of the lattice. The extracted thermal conductivities obtained for $ZrX_2$ show low thermal conductivities while quasi 1D-$ZrX_2$ exhibit thermal conductivity values that are similar to those reported for $ZrX_3$. Our results show that $ZrX_2$ materials exhibit exceptional thermal properties, making them candidates for thermoelectric device applications.

**ACKNOWLEDGEMENTS:**
This research was financially supported by King Abdulaziz City for Science and Technology (KACST) through the Center of Excellence for Green Nanotechnologies (CEGN), award 20132944. T.A.T and M.R.A would like to acknowledge the support of KACST cleanroom core lab facilities.

**METHODS:**
*Nanosheet Exfoliation and Confocal Raman Spectroscopy Measurements:*
ZrX2 nanosheets were obtained using micro-mechanical exfoliation technique. $ZrS_2$ and $ZrSe_2$ Crystals (2D- semiconductor) were exfoliated and deposited on $SiO_2$/Si substrate. Confocal Raman measurements (Renishaw) were taken using 532nm laser. To ensure repeatability of the observed Raman spectra, multiple exfoliation processes were carried out and different nanosheets were deposited on separate $SiO_2$/Si substrate.

Temperature dependence measurements were carried out using Linkam setup, where deposited nanosheets are placed inside a closed stage with optical window. The temperature of the stage was controlled using external equipment and Raman measurements were collected using 50X lens. For some room temperature measurements, where the optical window was not needed, 100X lens was used.

*Optothermal Energy Calculations:*
To compare spectra in figure S6 and S7, the optothermal energy is extracted. Here, laser power was measured using an optical power meter after each laser heating experiment. To calculate the delivered optothermal energy, the power density (*PD*) in units of (W/cm$^2$) is first calculated according to $PD = P_{laser}/(\pi r^2)$ where $r$ is the laser beam radius which depends on the objective lens used. Considering the exposure time ($T$), the optothermal energy delivered per unit area ($E$) is the integral of $PD$ taken over time in units of (J/cm$^2$). To obtained different optothermal energies, the laser power was varied along with the exposure time and the objective lens.

*Thermal Conductivity Model:*

The differential equation 2 can be solved by rewriting the heat equation in cylindrical coordinates, which can be expressed in a nonhomogeneous Bessel's equation

$$\frac{\partial^2 \theta}{\partial z^2} + \frac{1}{z}\frac{\partial \theta}{\partial z} - \theta = -\frac{q_0''}{g}\exp\left(-\frac{z^2}{z_0}\right) \qquad \text{Eq. 4}$$

Where:

$$\theta = \Delta T = T(r) - T_a, \qquad z = \sqrt{\frac{g}{kt}}r$$

The solution of this equation is:

$$\theta(z) = C_1 I_0(z) + C_2 K_0(z) + \theta_p(z) \qquad \text{Eq. 5}$$

Where the $I_0$ and $K_0$ are the zero-order modified Bessel functions of the first and second kind respectively. $\theta_p$ is the particular solution which can be found using the method of variational of parameters

$$\theta_p(z) = I_0(z)\int \frac{K_0(z)\frac{q_0''}{g}\exp\left(-\frac{z^2}{z_0^2}\right)}{-I_0(z)K_1(z) - K_0(z)I_1(z)}dz - K_0(z)\int \frac{I_0(z)\frac{q_0''}{g}\exp\left(-\frac{z^2}{z_0^2}\right)}{-I_0(z)K_1(z) - K_0(z)I_1(z)}dz \qquad \text{Eq. 6}$$

Finally, the average temperature rise is

$$\theta_m = \Delta T_m = \frac{\int \theta(r)\exp\left(-\frac{r^2}{r_0^2}\right)r\,dr}{\int \exp\left(-\frac{r^2}{r_0^2}\right)r\,dr} \qquad \text{Eq. 7}$$

These equations were solved numerically using MATLAB. The built-in modified Bessel's functions were used, and the trapezoidal integration method were carried out to solve the integral. The thermal conductivity and conductance values were found through an optimization that minimizes thermal conductivity based on matching the calculated $\theta_m$ with the experimentally measured $\Delta T_m$.

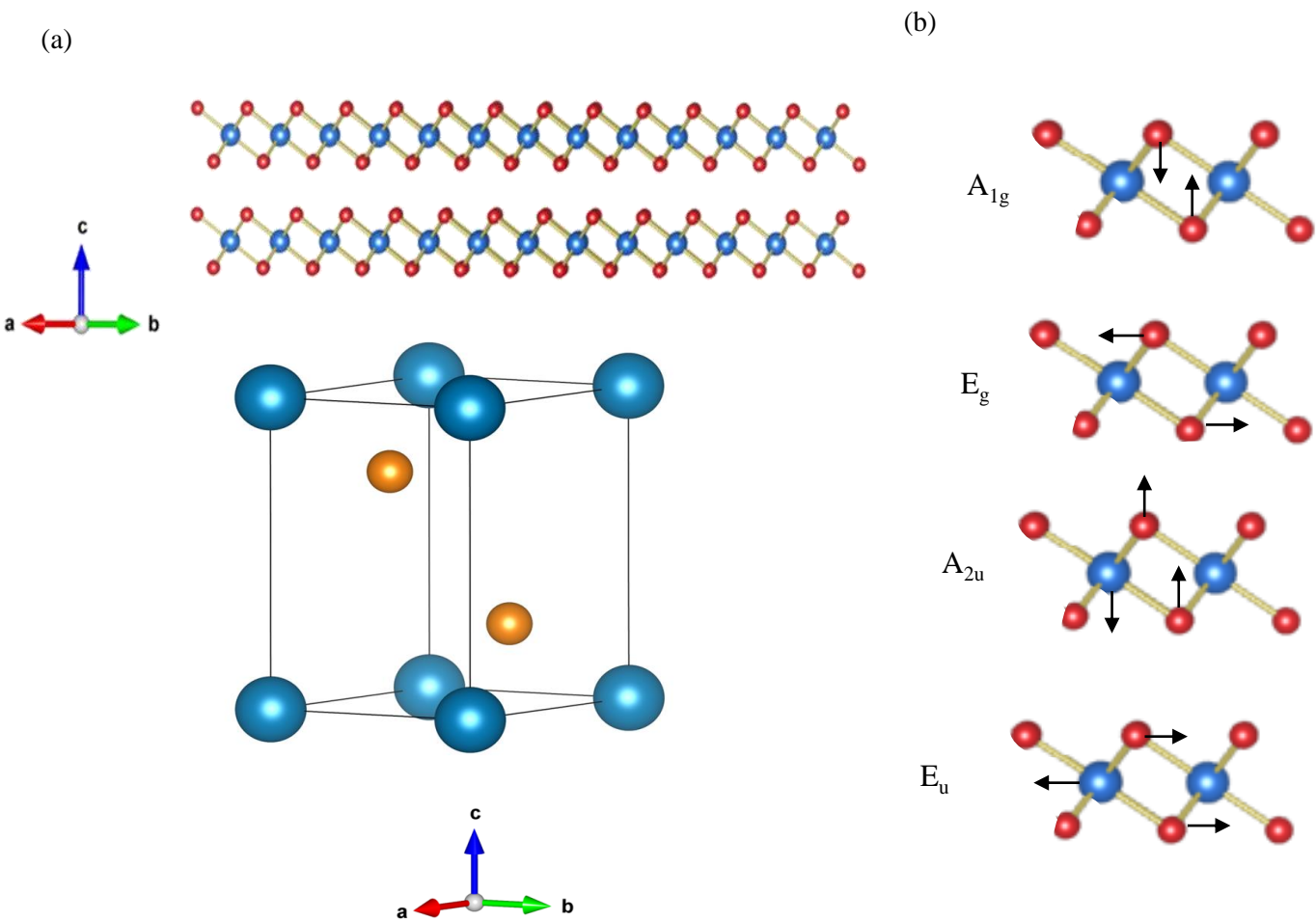

**Figure 1**. (a) Lattice structure and the primitive cell of ZrX2. In this structure, the angles of *ac* plane = *bc* plane = 90°, and *ab* plane = 120°. For $ZrS_2$, the parameters a = b = 3.663Å, and c = 5.827Å. While for $ZrSe_2$ a = b = 3.772Å, and c = 6.125Å. (b) Illustration of Raman-active modes measured for ZrX2 materials.

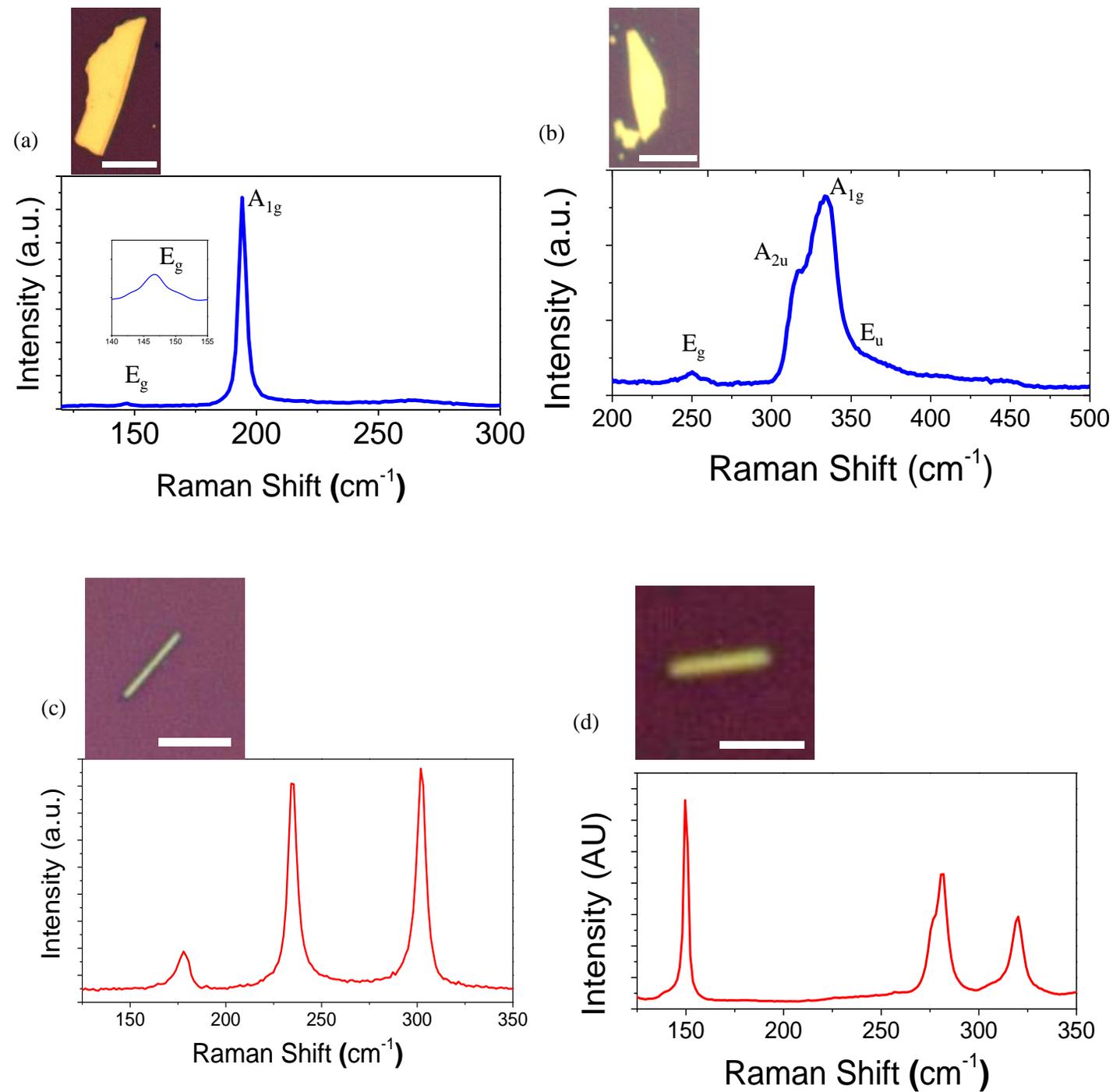

**Figure 2**. Raman spectra and optical images for (a) 2D-ZrSe$_2$, (b) 2D-ZrS$_2$, (c) quasi 1D-ZrSe$_2$, and (d) quasi 1D-ZrS$_2$.

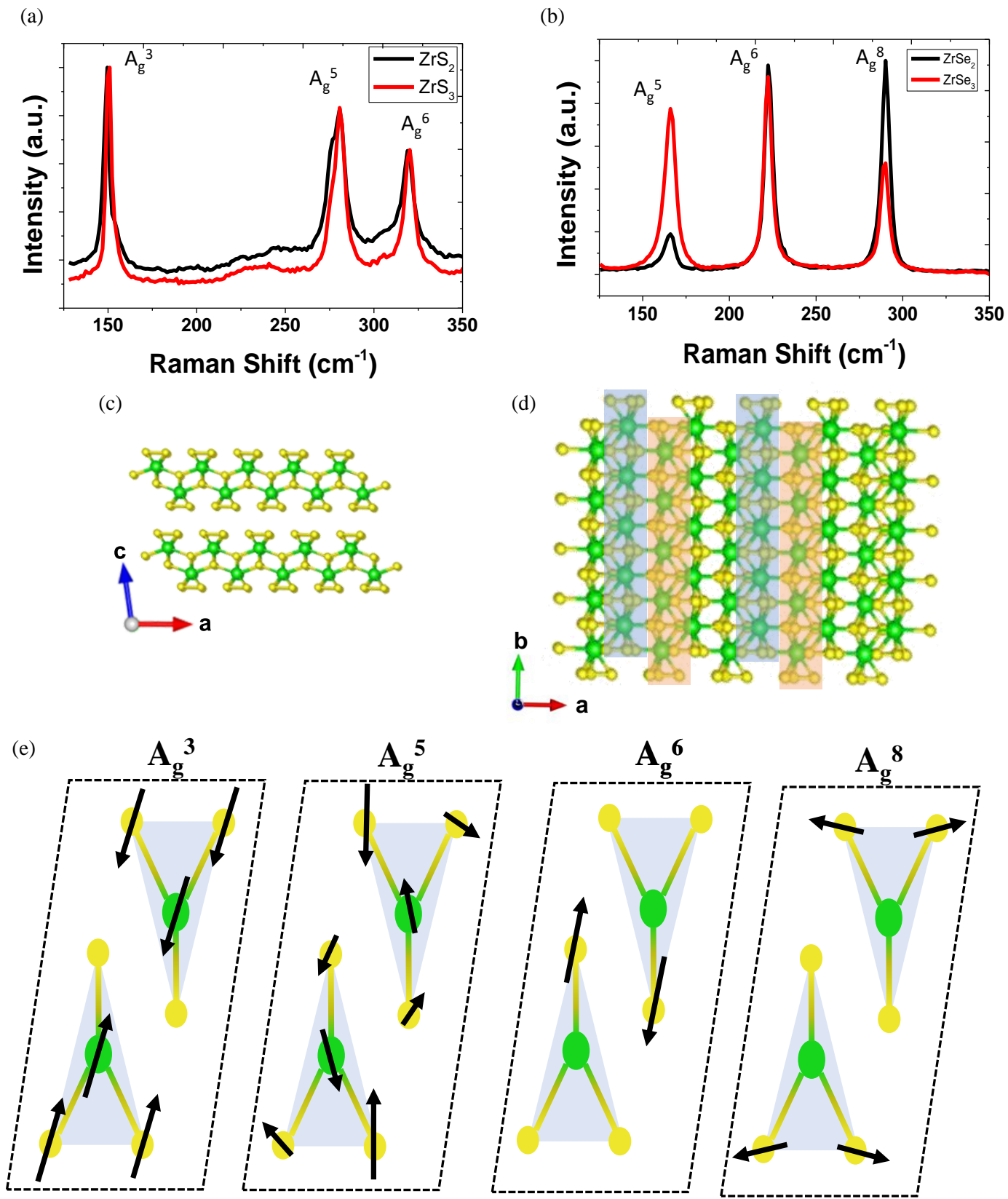

**Figure 3**. Raman spectra comparison between (a) quasi 1D-ZrS$_2$ and ZrS$_3$ (b) quasi 1D-ZrSe$_2$ and ZrSe$_3$. (c) Schematic diagram showing the cross-sectional viewed in *ac* plane of layered ZrX$_3$. (d) *ab* plane view of ZrX$_3$ structure with 1D-trigonal prismatic chain lattice highlighted in blue and pink. (e) Raman-active modes illustrated schematically for $A_g^3$, $A_g^5$, $A_g^6$, and $A_g^8$ modes.

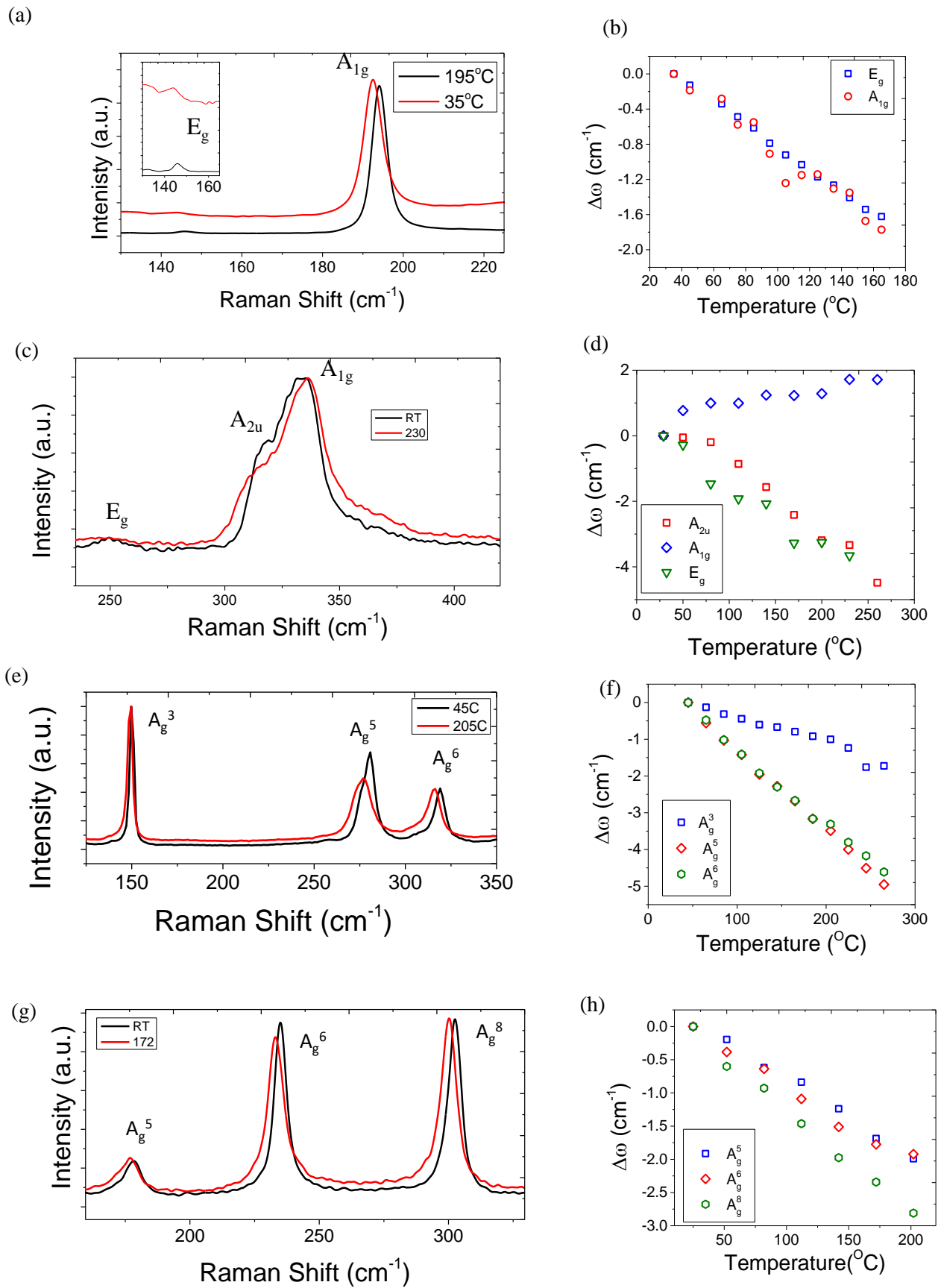

**Figure 4**. Raman spectra at 2 different temperatures for (a) 2D-$ZrSe_2$, (c) 2D-$ZrS_2$, (e) quasi 1D-$ZrS_2$, and (g) quasi 1D-$ZrSe_2$. The change in Raman shift ($\Delta\omega$) with varying temperature is plotted for (b) 2D-$ZrSe_2$, (d) 2D-$ZrS_2$, (f) quasi 1D-$ZrS_2$, and (h) quasi 1D-$ZrSe_2$.

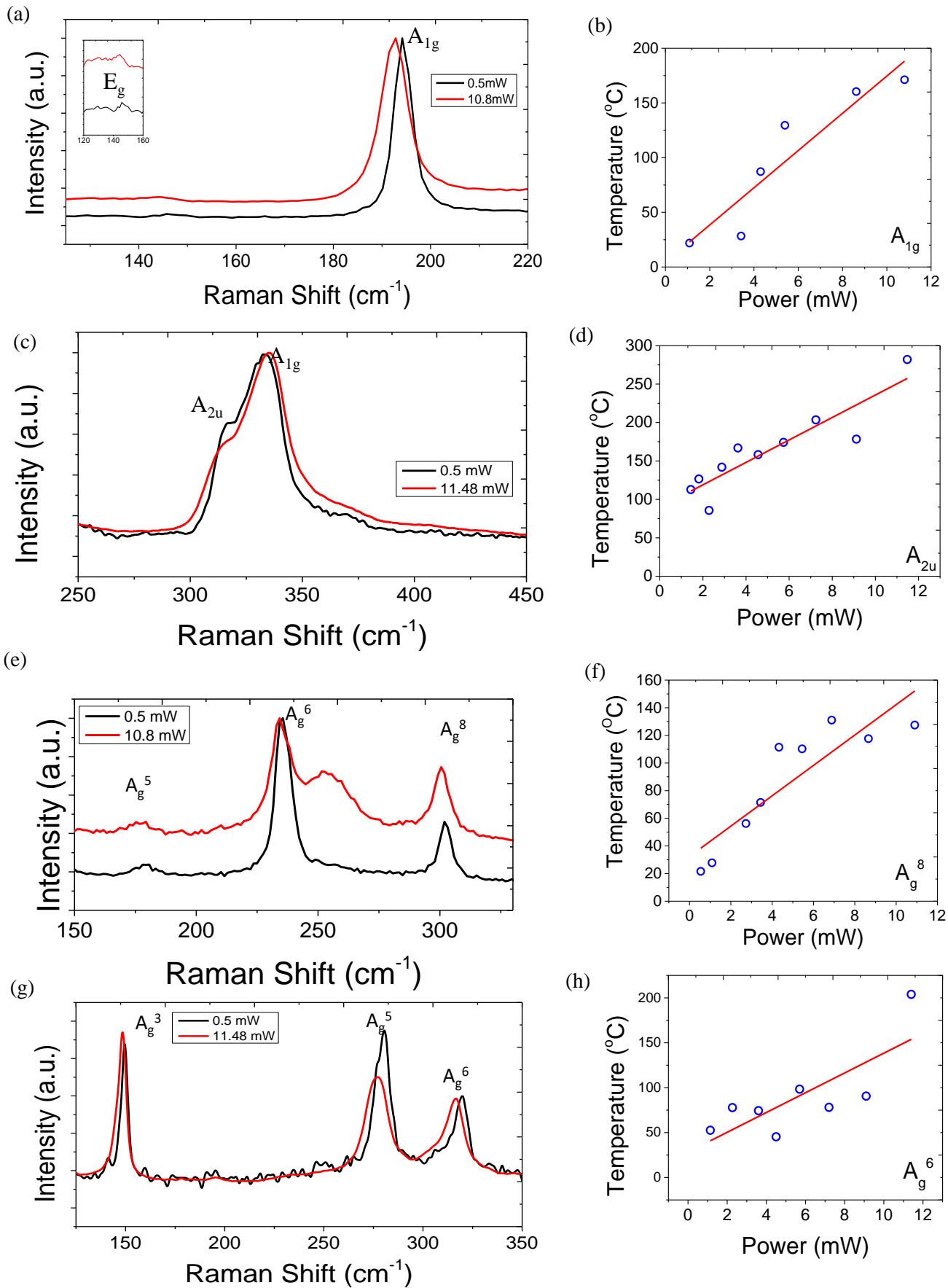

**Figure 5.** Raman spectra at 2 different optical heating powers for (a) 2D-ZrSe$_2$, (c) 2D-ZrS$_2$, (e) quasi 1D-ZrSe$_2$ and (g) quasi 1D-ZrS$_2$. The extracted temperature at different laser power is plotted for (b) 2D-ZrSe$_2$, (d) 2D-ZrS$_2$, (f) quasi 1D-ZrSe$_2$ and (h) quasi 1D-ZrS$_2$. The heating time for all materials was 1 second.

# Supporting Information for: Lattice Transformation from 2-D to Quasi 1-D and Phonon Properties of Exfoliated $ZrS_2$ and $ZrSe_2$

*Awsaf Alsulami[1], Majed Alharbi[1], Fadhel Alsaffar[1,2], Olaiyan Alolaiyan[1], Ghadeer Aljalham[1], Shahad Albawardi[1], Sarah Alsaggaf[1], Faisal Alamri[1], Thamer A. Tabbakh[3], and Moh R. Amer[1,4]\**

[1]Center of Excellence for Green Nanotechnologies,
Joint Centers of Excellence Program
King Abdulaziz City for Science and Technology, Riyadh, Saudi Arabia
[2]Department of Mechanical and Aerospace Engineering
University of California, Los Angeles, Los Angeles, CA, 90095
[3]National Center for Nanotechnology,
Materials Science Institute,
King Abdulaziz City for Science and Technology, Riyadh, Saudi Arabia
[4]Department of Electrical and Computer Engineering
University of California, Los Angeles, Los Angeles, CA, 90095

\* Please send all correspondence to mamer@seas.ucla.edu

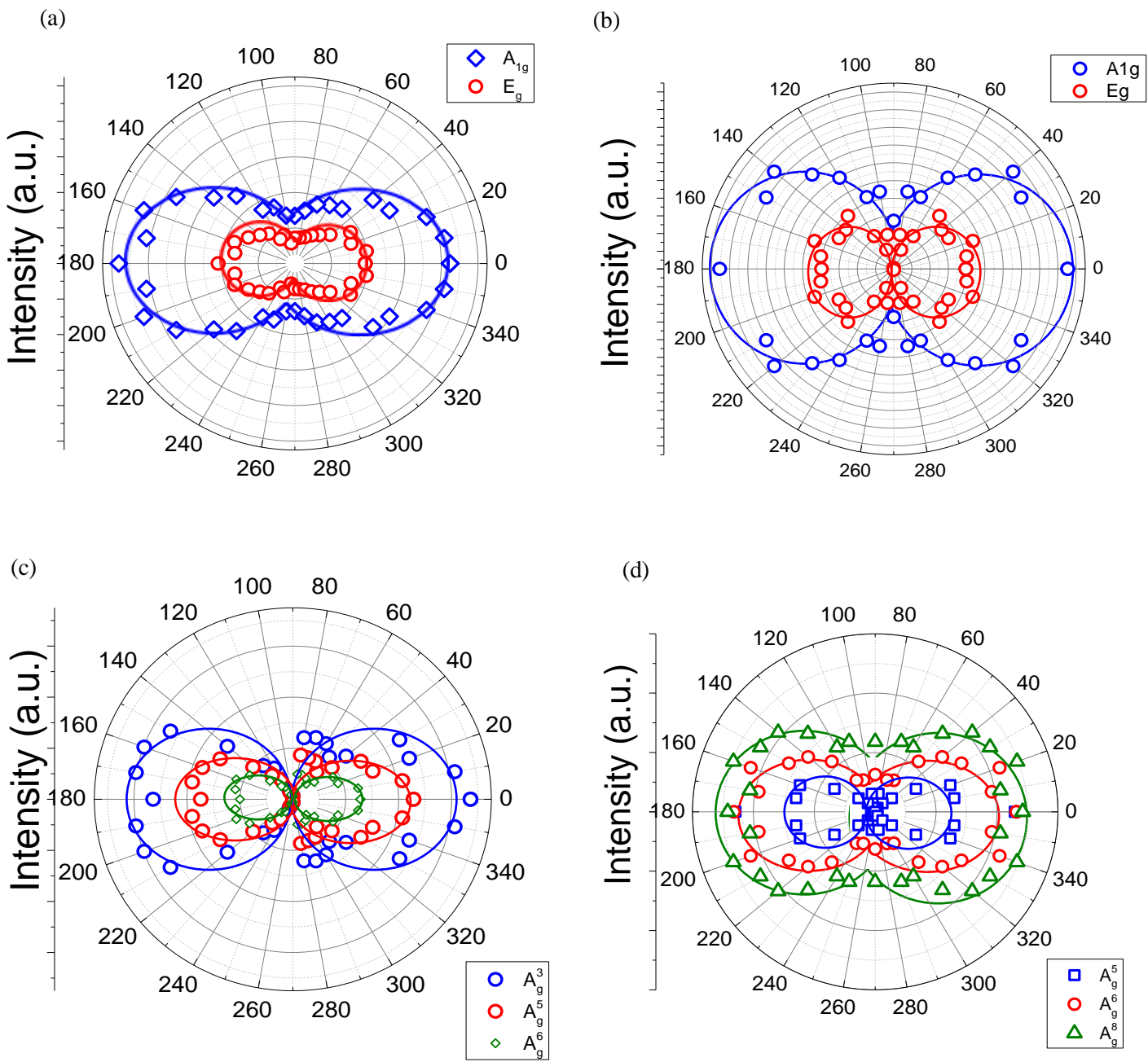

**Figure S1.** Polar plots of Raman intensity at different polarization angles for (a) 2D-ZrS$_2$ (b) 2D-ZrSe$_2$ (c) quasi 1D-ZrS$_2$ (d) quasi 1D-ZrSe$_2$.

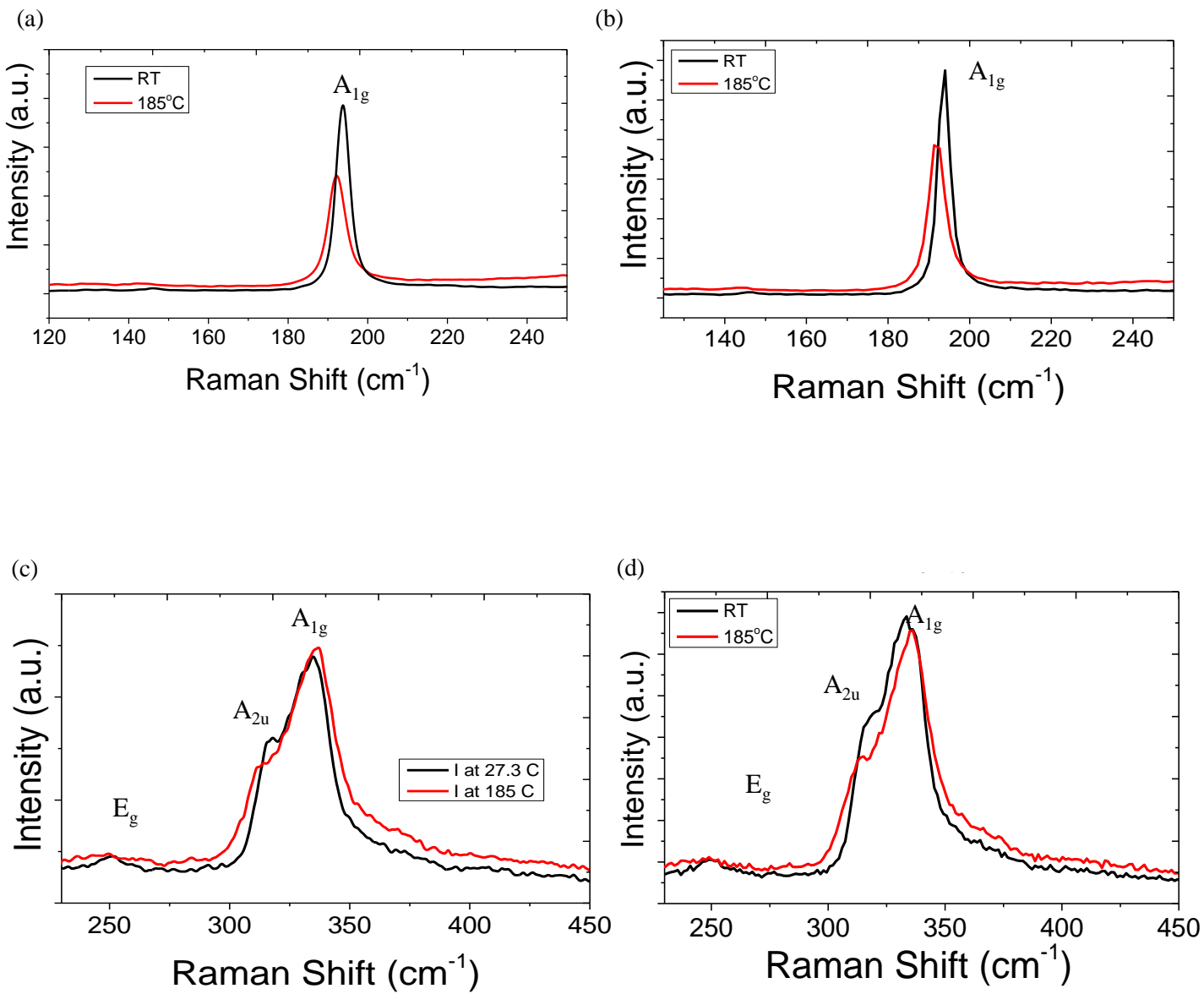

**Figure S2.** Raman spectra at different base temperatures for (a) sample 2 2D-ZrSe$_2$, (b) sample 3 2D-ZrSe$_2$, (c) sample 2 2D-ZrS$_2$, and (d) sample 3 2D-ZrS$_2$.

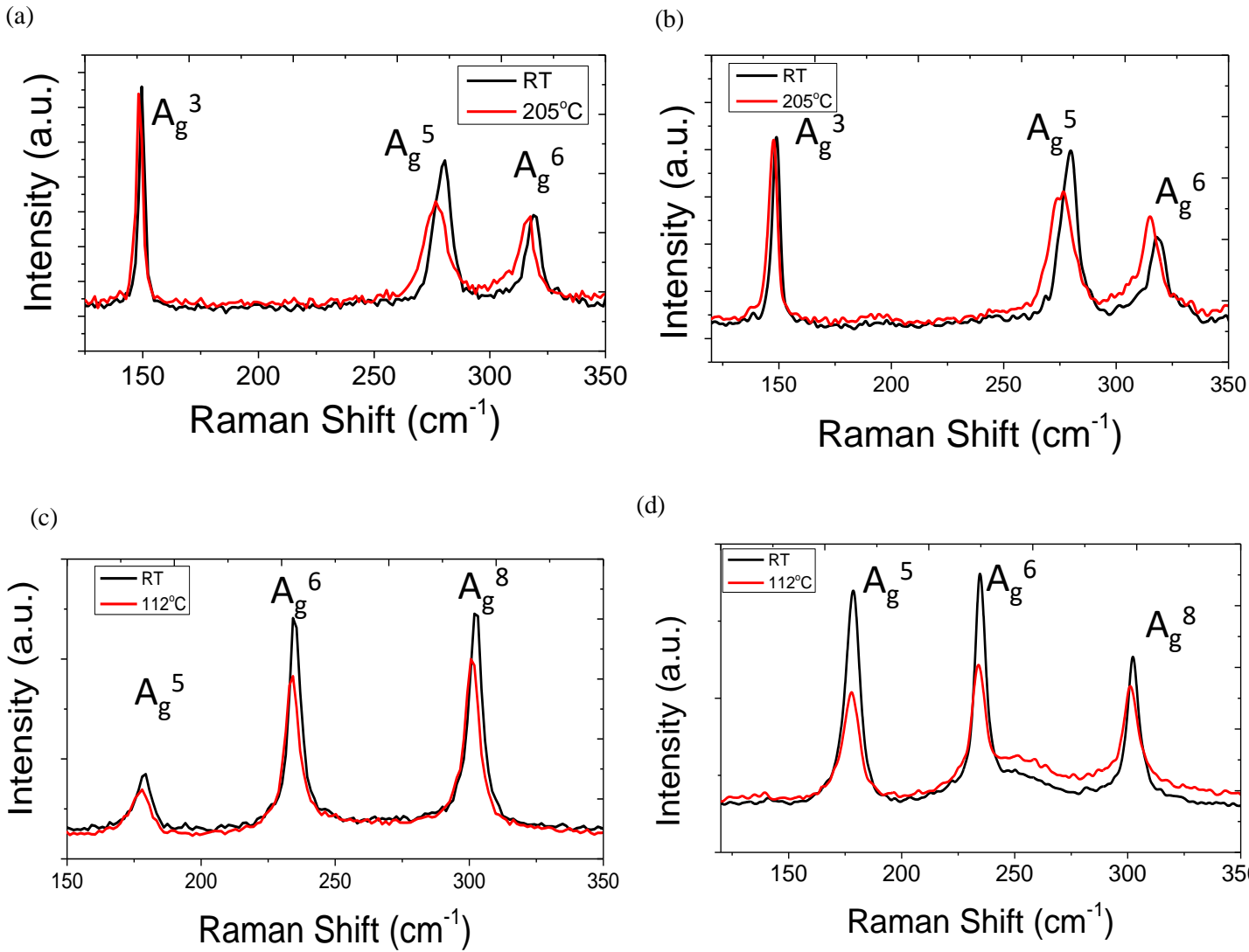

**Figure S3.** Raman spectra at different base temperatures for (a) sample 2 quasi 1D-$ZrS_2$, (b) sample 3 quasi 1D-$ZrS_2$, (c) sample 2 quasi 1D-$ZrSe_2$, and (d) sample 3 quasi 1D-$ZrSe_2$.

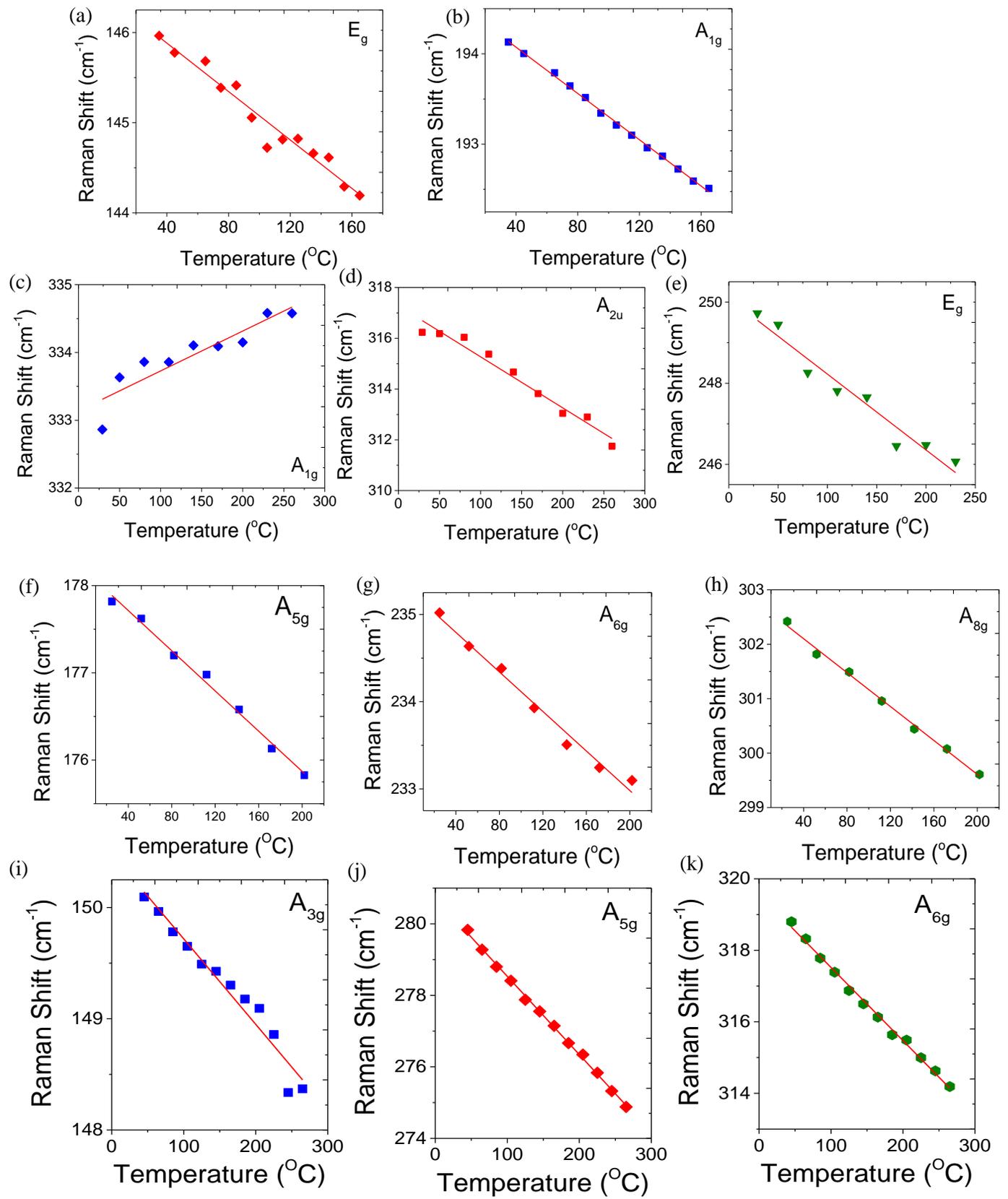

**Figure S4.** Raman shift vs. substrate temperature for (a) $E_g$ mode and (b) $A_{1g}$ mode for 2D-ZrSe$_2$, (c) $A_{1g}$ mode (d) $A_{2u}$ mode and (e) $E_g$ modes for 2D-ZrS$_2$, (f) $A_g^5$ mode (g) $A_g^6$ mode and (h) $A_g^8$ for quasi 1D-ZrSe$_2$, (i) $A_g^3$ mode (j) $A_g^5$ mode and (k) $A_g^6$ mode for quasi 1D-ZrS$_2$.

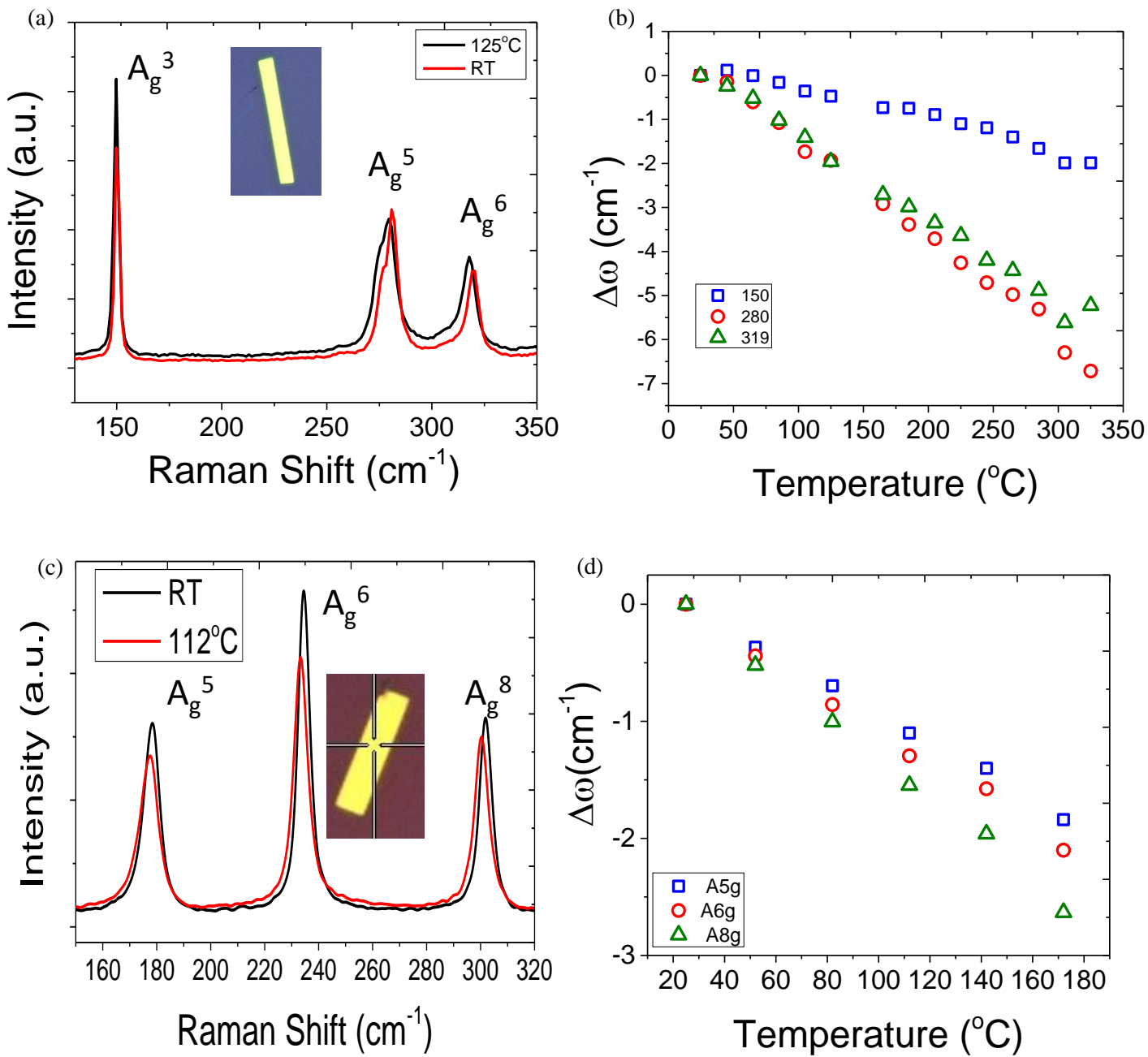

**Figure S5.** (a) Raman spectra at different substrate temperature and (b) the change in Raman shift (Δω) with varying temperature for ZrS$_3$. (c) Raman spectra at different substrate temperature and (b) the change in Raman shift (Δω) with varying temperature for ZrSe$_3$. Insets in (a) and (c) are nanosheet optical images.

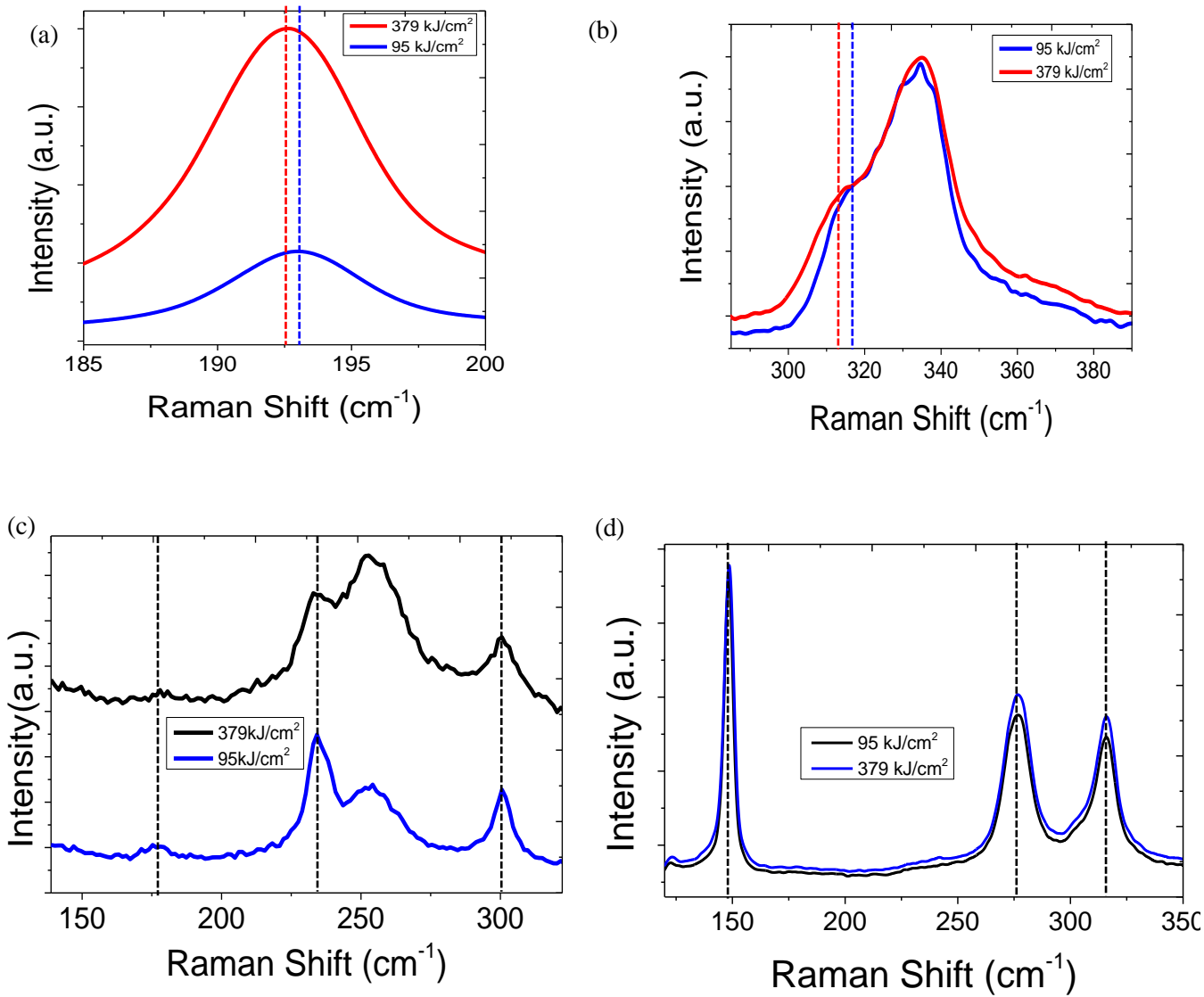

**Figure S6.** (a) Raman spectra at different laser heating energy (kJ/cm$^2$) for (a) 2D-ZrSe$_2$ (b) 2D-ZrS$_2$ (c) quasi 1D-ZrSe$_2$ (d) quasi 1D-ZrS$_2$. The dashed line is to highlight and compare the downshift at different laser heating energies. Notice for A$_{1g}$ of 2D-ZrS$_2$ in (b) the anomalous upshift is almost identical which stems from the weak dependence of this mode on temperature. Moreover, the additional mode that occurs at ~254cm$^{-1}$ is due to oxidation of the nanosheet with high laser power.

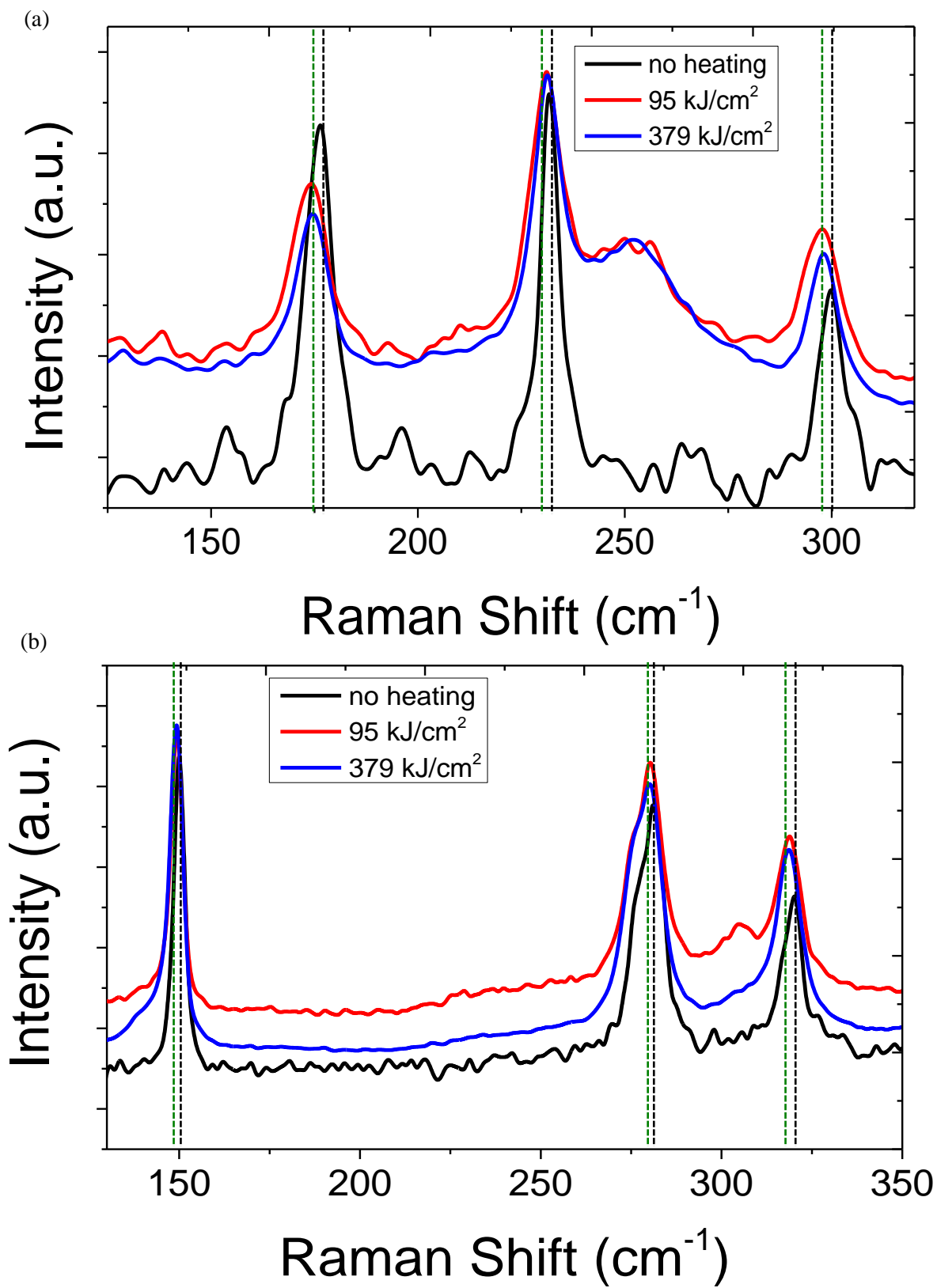

**Figure S7.** (a) Raman spectra at different laser heating energy (kJ/cm$^2$) for (a) ZrSe$_3$ and (b) ZrS$_3$. Notice the downshift for all modes is identical regardless of the laser heating energy.

(a)

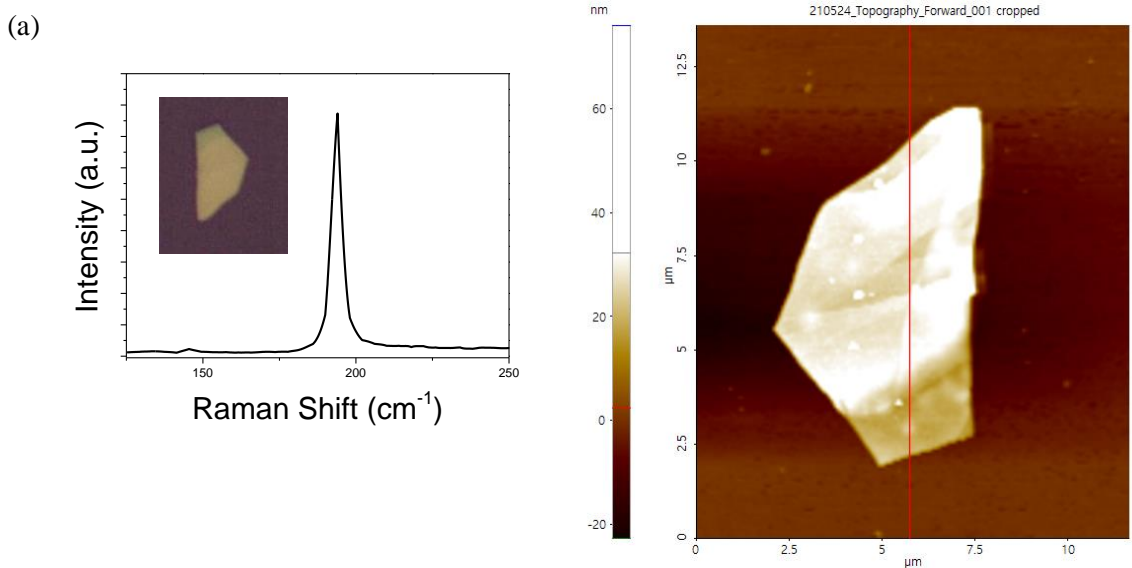

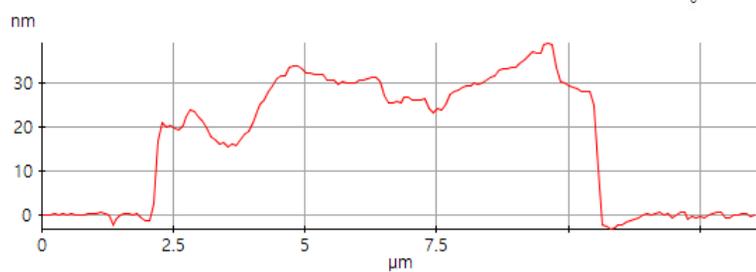

(b)

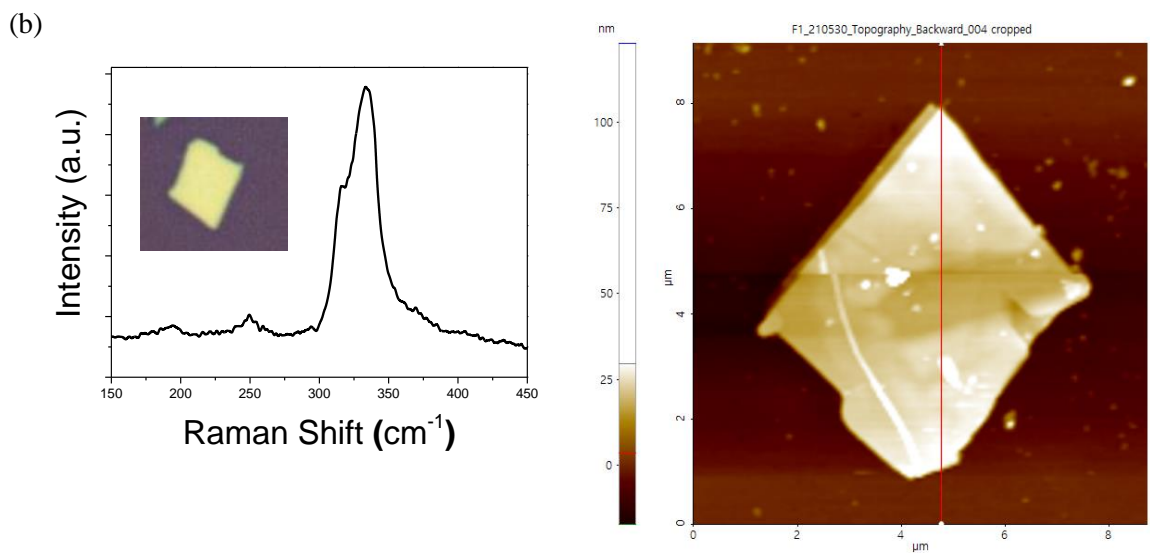

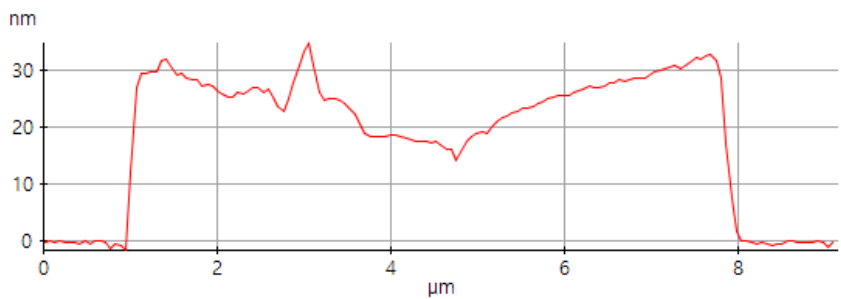

**Figure S8.** (a) Raman spectrum, optical image, and AFM measurements of (a) 2D-$ZrSe_2$ and (b) 2D-$ZrS_2$.

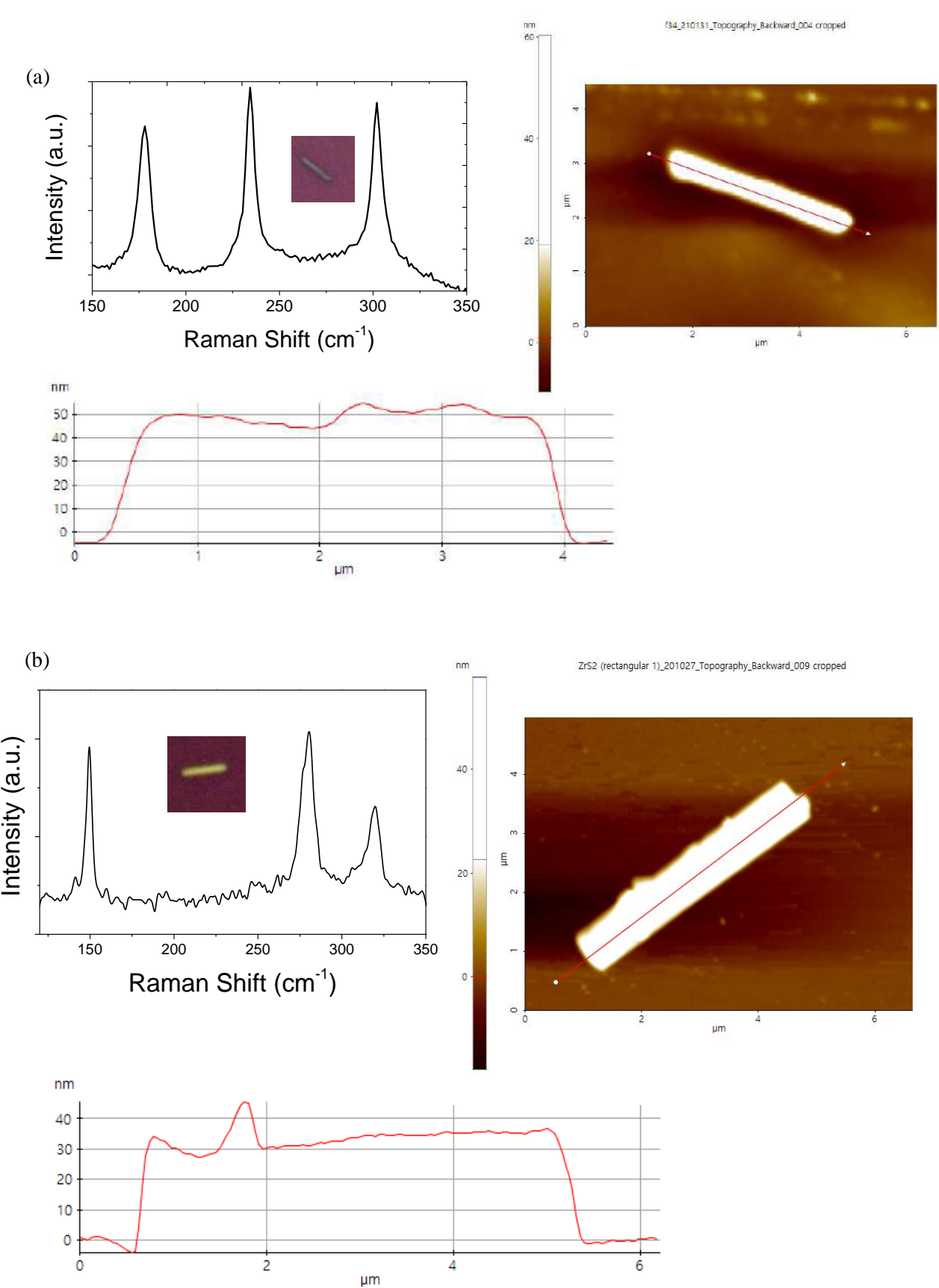

**Figure S9.** (a) Raman spectrum, optical image, and AFM measurements of (a) quasi 1D-ZrSe$_2$ and (b) quasi 1D-ZrS$_2$.